**[Title]** Effect of forename string on author name disambiguation

**[Authors]** Jinseok Kim and Jenna Kim

**[Author Information]**

1. Jinseok Kim (Corresponding Author)

Institute for Research on Innovation & Science, Survey Research Center, Institute for Social Research
University of Michigan
330 Packard Street, Ann Arbor, MI U.S.A. 48104
jinseokk@umich.edu

2. Jenna Kim

School of Information Sciences
University of Illinois at Urbana-Champaign
501 E. Daniel Street, Champaign, IL U.S.A. 61820
jkim682@illinois.edu

Abstract

In author name disambiguation, author forenames are used to decide which name instances are disambiguated together and how much they are likely to refer to the same author. Despite such a crucial role of forenames, their effect on the performances of heuristic (string matching) and algorithmic disambiguation is not well understood. This study assesses the contributions of forenames in author name disambiguation using multiple labeled datasets under varying ratios and lengths of full forenames, reflecting real-world scenarios in which an author is represented by forename variants (synonym) and some authors share the same forenames (homonym). Results show that increasing the ratios of full forenames improves substantially the performances of both heuristic and machine-learning-based disambiguation. Performance gains by algorithmic disambiguation are pronounced when many forenames are initialized or homonym is prevalent. As the ratios of full forenames increase, however, they become marginal compared to the performances by string matching. Using a small portion of forename strings does not reduce much the performances of both heuristic and algorithmic disambiguation compared to using full-length strings. These findings provide practical suggestions such as restoring initialized forenames into a full-string format via record linkage for improved disambiguation performances.

# Introduction

In author name disambiguation, author names are the source of ambiguity: same names can refer to different authors (homonym), while an author can be represented by different name variants (synonym). At the same time, author names provide clues to solve the ambiguity problem. In disambiguation research, specifically, author names are used to decide which pairs of instances will be disambiguated together (blocking). In addition, many disambiguation heuristics and algorithms rely on author name similarity to reach a decision of match or non-match among blocked name instances. During this process, name instances that share full forenames are often assumed to represent the same author (e.g., Cota, Ferreira, Nascimento, Gonçalves, & Laender, 2010; Ferreira, Veloso, Gonçalves, & Laender, 2014; Kim & Diesner, 2016; Liben-Nowell & Kleinberg, 2007; Onodera et al., 2011; Xie, Ouyang, Li, Dong, & Yi, 2018).

Such presumed importance of forenames has, however, been insufficiently evaluated. Many aforesaid studies have assumed forename matching between name instances as predictive of the same author identity without empirical validation. A few studies have attempted to quantify the forename contributions in heuristic and algorithmic disambiguation (e.g., Louppe, Al-Natsheh, Susik, & Maguire, 2016; Müller, Reitz, & Roy, 2017; Torvik & Smalheiser, 2009), but not provided knowledge about how the contribution of forename can change in comparison with other features such as coauthorship, title, and venue under different ambiguity settings like varying full forename ratios. As a result, it is not well known how many performance gains can be obtained by using forenames in blocking name instances and calculating similarity among them. This deficiency of knowledge can be an obstacle to improving future disambiguation efforts.

The present study tries to address the deficiency by assessing the contributions of forenames in both heuristic and algorithmic disambiguation using various labeled datasets. For this purpose, name ambiguity settings are simulated in two ways. First, the ratios of full forenames in each dataset are changed from 0% (all initialized) to 100%. Second, forename strings are stripped into 1 through 10 characters. These settings reflect the real-world scenario in which an author can be represented by different forename variants (synonym) and two or more authors share the same forenames (homonym). Next, the performances of simple string-based matching (heuristic) and machine-learning-based (algorithmic) disambiguation are evaluated under each setting. In addition, for algorithmic disambiguation, the forename's feature importance in relation to other features is measured by impurity change of Random Forest. Findings from this study will help us better understand the effectiveness of forenames in author name disambiguation and provide practical implications for improving future disambiguation studies. In the following section, related work is discussed to contextualize the study.

# Related Work

A common use of forenames in author name disambiguation is to collect name instances to be compared for disambiguation into a block (blocking). Many studies have blocked name instances that match on a full surname and a forename's first initial, while several others have also used the full forename + surname for blocking (Cota et al., 2010; Liu, Li, Huang, & Fang, 2015; Wang, Tang, Cheng, & Yu, 2011; Zhu et al., 2018). Another usage of forenames is to calculate the forename similarity between a pair of name instances in a block, whose match or non-match is decided heuristically (e.g., match if above a certain threshold) (e.g., Kim & Diesner, 2016; Martin, Ball, Karrer, & Newman, 2013) or by machine learning algorithms trained on the similarity scores of instance pairs labeled 'match' or 'non-match' (e.g., Louppe et al., 2016; Song, Kim, & Kim, 2015). Some algorithmic disambiguation studies have investigated which feature contributes most to disambiguation results. Coauthorship has been reported to be most effective (Liu et al., 2014; Onodera et al., 2011; Torvik, Weeber, Swanson, & Smalheiser, 2005; Wang et al., 2011). Sometimes, affiliation information has been found to be more important than coauthorship (Song et al., 2015; Wu & Ding, 2013).

Although various features have been evaluated for their impacts on disambiguation, the effectiveness of forenames has been insufficiently investigated. Many studies have assumed that two name instances belong to the same author if they share a forename (and surname) or assigned to the pair a higher similarity score than when the pair is similar over other features (e.g., Cota et al., 2010; Ferreira et al., 2014; Han, Giles, Zha, Li, & Tsioutsiouliklis, 2004; Kim, 2018; Kim & Diesner, 2016; Onodera et al., 2011; Xie et al., 2018). However, those studies have not validated such presumed effectiveness of forename strings on labeled data, especially, in comparison with other features.

A small number of studies have tested how much forename string affects author name disambiguation. For example, Han et al. (2004) stripped all forenames leaving the first three characters left (e.g., 'Mark E. J. Newman' → 'Mar Newman') and found that such stripping improved greatly machine learning performance from a baseline in which all forenames were initialized (e.g., 'Mark Newman' → 'M Newman'). Another example is Louppe et al. (2016), which found that full forenames improved marginally disambiguation results in a dataset where roughly 40% of name instances have full forenames. Recently, forename string matching has been shown highly accurate in disambiguating names (Backes, 2018; Müller et al., 2017). These studies, however, analyzed a single dataset or a specific name ambiguity condition. So, they cannot tell us how their findings hold true for datasets with different forename conditions such as varying ratios and lengths of full forenames. Furthermore, forename's contributions in author name disambiguation have rarely been compared across heuristic and algorithmic approaches.

As such, the effectiveness of forenames in author name disambiguation is an under-researched topic. Studying the topic has a potential to provide actionable insights into improving author name disambiguation. Methodologically, a proper understanding of forename's impact on disambiguation can lead to developing simpler but more scalable disambiguation approaches based on forenames than many current methods mining multiple features causing a high computational burden. For example, drawing on the finding that coauthorship plays an important role in distinguishing authors, several coauthorship-focused algorithms have been proposed (e.g., Kang et al., 2009; Shin, Kim, Choi, & Kim, 2014) and implemented on a digital library scale (e.g., Reitz & Hoffmann, 2013).

Another reason to pay attention to forenames in author name disambiguation is that full forenames in digital libraries keep increasing in number. Figure 1 shows yearly and cumulative ratios of name instances with full forenames (circles), as recorded in two bibliographic datasets – MEDLINE[1] (87.4 million name instances) and Microsoft Academic Graph[2] (MAG; 294.6 million) – over the 1950~2015 period. The yearly ratios increase up to 90% in MEDLINLE and 80% in MAG, while the cumulative ratios up to 46% and to 76% in 2015. The abrupt change around 2002 in MEDLINE is due to the indexing policy change by the National Library of Medicine to record author forenames in a full-string format, if available. Contrary to the trends, many disambiguation studies have focused on solving ambiguity problems in which initialized forenames are dominant (Müller et al., 2017) or full forenames are available for most name instances (see Table 1 below), constraining the applicability of their models and findings to disambiguating names in libraries in which full forenames are becoming dominant but not complete for all name instances.

---

[1] ftp://ftp.ncbi.nlm.nih.gov/pubmed/baseline. This study used the 2016 baseline.
[2] https://www.openacademic.ai/oag/. This study used the 2016 release.

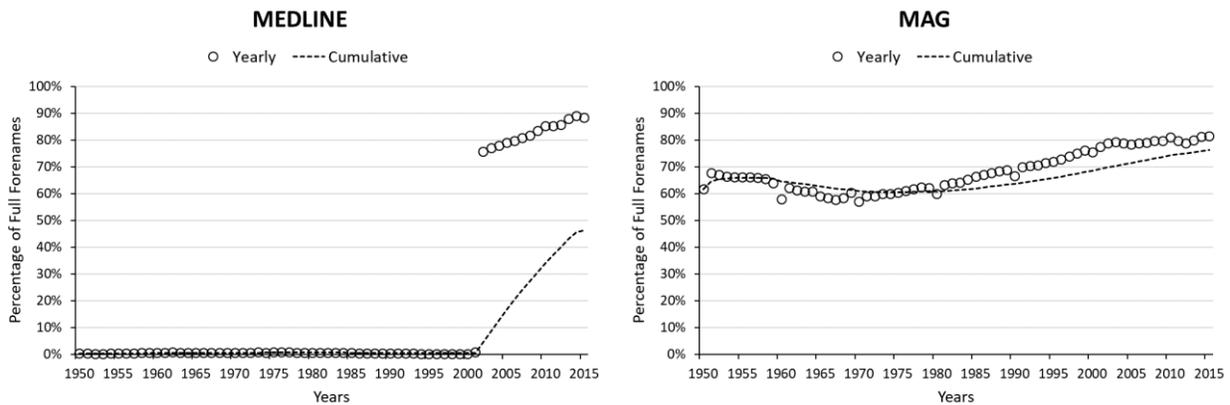

*Figure 1: Yearly and Cumulative Percentage of Full Forenames in Two Large-Scale Bibliographic Databases*

For these reasons, the goal of this study is to evaluate the effectiveness of forenames in author name disambiguation by measuring the performance changes of heuristic and algorithmic methods in three labeled datasets under varying ratios of full forenames and lengths of forename strings. In each setting, this study quantifies relative contributions of forenames in algorithmic disambiguation in comparison with other features using the feature importance measurement in Random Forest. In the following section, datasets, machine learning set-ups, and evaluation measures are explained in details.

## Methods

Datasets

The impact of forename string on author name disambiguation is measured with four labeled datasets that have been widely used in many studies, separately or jointly (e.g., Cota et al., 2010; Ferreira et al., 2014; Kim & Kim, 2018; Momeni & Mayr, 2016; Müller et al., 2017; Pereira et al., 2009; Santana, Gonçalves, Laender, & Ferreira, 2017; Shin et al., 2014; Wu, Li, Pei, & He, 2014; Zhu et al., 2018).

PENN[3]: Labeled for Han et al. (2004) by researchers at the Pennsylvania State University, this dataset was originally comprised of 8,453 name instances with their coauthorship, paper title, and venue information. As its original version contained duplication and labeling errors, several studies modified the dataset before use (e.g., Cota et al., 2010; Kim & Kim, 2018; Santana et al., 2017; Shin et al., 2014). This study re-uses one of recent revisions by Kim (2018) in which 5,018 name instances and their associated metadata are linked to DBLP records after de-duplication and verification of correctness[4].

KISTI[5]: This consists of 41,673 name instances of 6,921 authors and their associated information such as coauthorship, paper title, publication year, and venue, which were extracted from DBLP (Kang, Kim, Lee, Jung, & You, 2011). Name ambiguity was resolved by researchers at the Korea Institute of Science Technology & Information (KISTI) and Kyungsung University through aggregating Google search results and manual disambiguation outcomes.

---

[3] http://clgiles.ist.psu.edu/data/nameset_author-disamb.tar.zip
[4] https://doi.org/10.6084/m9.figshare.6840281.v2
[5] http://www.lbd.dcc.ufmg.br/lbd/collections/disambiguation/DBLP.tar.gz/at_download/file

AMINER[6]: This dataset was labeled for Wang et al. (2011), and updated later to train and test disambiguation algorithms for AMiner, a digital library aggregating publication records from several computing digital libraries (Tang et al., 2008). The dataset contains 7,528 homonym instances (of 1,546 authors) which are associated with coauthorship, affiliation, paper title, publication year, and venue.

GESIS[7]: Created in the GESIS – Leibniz Institute for the Social Sciences, this labeled dataset is a collection of 5,408 authors who have homonyms and their publication records in DBLP (Momeni & Mayr, 2016). For this study, especially, 29,965 name instances of 2,580 authors in the Evaluation Set are used. In addition, the original DBLP records (2015 May version) are linked to the selected name instances to associate them with title and venue information which is not recorded in GESIS.

Table 1 reports the number of name instances in each dataset and the ratios of full forenames that are counted from author names to be disambiguated and their coauthors. If a name instance has no forename, 'Null' is assigned. According to the table, about 96~97% of name instances in each dataset are recorded with one or more full forenames[8]. Reflecting a real world scenario in which forenames are recorded in an initialized-string (e.g., C. Brown) or a full-string (Charles Brown) format, this study differentiates the ratios of full forenames in each dataset from zero to 100 percent of name instances that have full forenames with a 10% increment (e.g., 0%, 10% … 100%). For example, if 10% of name instances are chosen to be with full forenames, the forenames of the remaining name instances (90%) are initialized. During this process, some name instances of an author may be initialized, while others are not. This selection of 10~90% of name instances in each dataset is conducted randomly and repeated for 10 times.

*Table 1: Summary of Labeled Datasets*

| Data | Year Created | No. of Name Instances for Disambiguation | No. of Name Instances including Coauthors | Forename String Type (%) | | |
|---|---|---|---|---|---|---|
| | | | | Null | All Initialized | One or More Full |
| PENN | 2004 | 5,018 | 12,730 | 9 (0.07%) | 454 (3.57%) | 12,267 (96.36%) |
| KISTI | 2010 | 41,673 | 116,236 | 12 (0.01%) | 2,979 (2.56%) | 113,245 (97.43%) |
| AMINER | 2011 | 7,528 | 26,560 | 16 (0.06%) | 1,127 (4.24%) | 25,417 (95.70%) |
| GESIS | 2015 | 29,965 | 112,318 | 20 (0.02%) | 1,183 (1.05%) | 111,116 (98.92%) |

Machine Learning Setups

Blocking: Only name instances that match on a full surname and the first initial of a forename are compared pairwisely for disambiguation. This blocking method has been commonly used in disambiguation studies because it can improve computational efficiency by reducing the number of comparable pairs among name instances. For example, if we want to disambiguate 1,000 name instances, we need to compare 499,500 pairs without blocking, but 49,500 pairs ($\approx$ 1/10) with 10 blocks each containing 100 comparable name instances. Blocking can miss instance pairs that refer to the same author but belong to different blocks. Misidentification of authors due to this problem is, however, acceptable with, for example, 1.2% of errors (Torvik & Smalheiser, 2009).

---

[6] http://arnetminer.org/lab-datasets/disambiguation/rich-author-disambiguation-data.zip
[7] http://dx.doi.org/10.7802/1234
[8] Surname and forename of an author name instance are distinguished in PENN, KISTI, and AMINER, but not in GESIS. After each name instance in GESIS is separated by spaces, its last token (unless it is one-character long or contains suffices like Jr, II, etc.) is assumed as a surname and remaining tokens forenames.

Pairwise Similarity Calculation: Four features – forename, coauthor, title, and venue – are used for similarity calculation between name instances because they are common to all our datasets and have been used in many disambiguation studies (Schulz, 2016; Song et al., 2015). Two name instances in a block are compared for similarity over these four features as follows. Each name string is lower-cased, converted into ASCII format, and segmented into an array of 2~4-grams, following several studies (Han, Xu, Zha, & Giles, 2005; Kim & Kim, 2018; Kim, Kim, & Owen-Smith, 2019; Louppe et al., 2016; Treeratpituk & Giles, 2009). For example, 'Mark' is converted into a list of 'ma,' 'ar,' 'rk,' 'mar,' 'ark,' and 'mark.' Then, a cosine similarity of the term frequency (TF) between the 2~4-gram lists of two name instances is calculated as a forename similarity score for the instance pair. The series of character conversion, *n*-gram segmentation, and TF-based cosine similarity calculation are applied to other features. For title words, stop-words[9] are filtered and each remaining word is stemmed by the Porter's algorithm (Porter, 1980)[10] before similarity calculation.

Algorithmic Model Learning: Both blocking and feature similarity calculation are conducted on name instances in two subsets – training and test (validation) sets – that are randomly split (50%-50%) from each dataset with different ratios (0%~100%) of full forenames. The similarity scores among name instance pairs over four features are fed into five algorithms – Random Forest, Naïve Bayes, Logistic Regression, Support Vector Machine, and Gradient Boosting Trees – that have been used as baseline or the best classifiers in many disambiguation studies (e.g., Han et al., 2004, 2005; Kim et al., 2019; Kim, Sefid, Weinberg, & Giles, 2018; Levin, Krawczyk, Bethard, & Jurafsky, 2012; Louppe et al., 2016; Santana et al., 2017; Song et al., 2015; Torvik & Smalheiser, 2009; Treeratpituk & Giles, 2009; Wang et al., 2012). Each algorithm learns disambiguation patterns using the feature similarity scores generated above for name instance pairs labeled as match (positive pairs) or non-match (negative pairs). To reduce computational burden from running SVM, 10~20% of pairs are randomly selected for training.

Prediction and Clustering: Based on disambiguation models learned from training data, each algorithm assigns a 'match' probability score (ranging from 0 to 1) to instance pairs in test data. This prediction score is used as a similarity distance between the pair to be fed into a hierarchical agglomerative clustering algorithm, which groups name instances into a cluster if their distances are above a certain threshold. Following previous studies (Kim & Kim, 2018; Levin et al., 2012; Liu et al., 2014; Louppe et al., 2016; Torvik & Smalheiser, 2009), this threshold is decided by trying various distance values between 0 and 1 in each block and choosing one that produces the best clustering result (measured by B-Cubed F1 explained below) for the block[11]. Meanwhile, to evaluate how a heuristic performs in comparison with algorithmic disambiguation, name instances in test data that match on all available name (surname + forename) strings are assumed to refer to the same author. This string-based matching of authors has been dominantly used as a heuristic disambiguation method in bibliometrics (for details, see Kim and Diesner (2016)).

Performance Evaluation

---

[9] https://github.com/stanfordnlp/CoreNLP/blob/master/data/edu/stanford/nlp/patterns/surface/stopwords.txt
[10] https://tartarus.org/martin/PorterStemmer/
[11] All classifiers were implemented using Scikit-learn packages. An optimal threshold for a hierarchical agglomerative clustering was decided by implementing the code at https://github.com/glouppe/paper-author-disambiguation (Louppe et al., 2016).

As a result of algorithmic and heuristic disambiguation above, name instances decided to refer to the same author are gathered into a cluster. These clusters predicted by a disambiguation method from a test dataset are compared to truth clusters that are generated from labels of name instances in the same test dataset. Disambiguation performances are evaluated by B-Cubed (B³), following previous studies (Delgado, Martínez, Montalvo, & Fresno, 2017; Kim & Kim, 2018; Levin et al., 2012; Louppe et al., 2016; Momeni & Mayr, 2016; Müller et al., 2017; Qian, Zheng, Sakai, Ye, & Liu, 2015). This measure consists of three parts – B³ Recall ($R$), B³ Precision ($P$), and B³ F ($F$) – defined as follows (Levin et al., 2012):

$$R = \frac{1}{N} \sum_{t \in T} \frac{|P(t) \cap T(t)|}{|T(t)|} \quad (1)$$

$$P = \frac{1}{N} \sum_{t \in T} \frac{|P(t) \cap T(t)|}{|P(t)|} \quad (2)$$

$$F = \frac{2 \times R \times P}{R + P} \quad (3)$$

Here, $t$ is a name instance in truth clusters $T$. $N$ is the number of all name instances in truth clusters ($T$). $T(t)$ is a truth cluster that contains a name instance $t$, while $P(t)$ a predicted cluster that contains the name instance $t$. B³ measure is calculated using the fast algorithm proposed in Kim (2019).

Feature Importance Measure

To assess each feature's contribution in algorithmic disambiguation, the feature importance measure in Random Forest is used because it provides more stable results than the sequential-forward/backward-elimination method (Guyon & Elisseeff, 2003; Saeys, Abeel, & Van de Peer, 2008). When implemented on training data, Random Forest calculates how much the tree nodes that use a specific feature – one of forename, coauthor, title, and venue - reduce on average impurity across all trees in the forest[12]. If a feature decreases mean impurity more than other features, the feature is considered more important than others. Mean impurity scores reduced by each feature are scaled so that the sum of all features' scores equals one, and each feature is assigned the rescaled score as its feature importance score.

Name Ambiguity Type

To see how the change of full forename ratios affects the composition of name ambiguity in each dataset, four name ambiguity types are counted per instance pair. Four types include: homonym (same name strings representing different authors), synonym (different name strings representing same authors), SN-SA (same name strings representing same authors), and DN-DA (different name strings representing different authors) (Louppe et al., 2016). Note that ambiguity type is decided for an instance pair because the feature similarity calculation in algorithmic disambiguation is conducted at an instance-pair level within a block in this paper.

---

[12] 500 trees were used after grid search for Random Forest. Gini Impurity was chosen for split quality measure. Other notable settings include: L2 Regularization with class weight = 1 for Logistic Regression, Gaussian Naïve Bayes with maximum likelihood estimator for Naïve Bayes, Linear Kernel for SVM, and 500 estimators (max depth = 9; learning rate = 0.125) for Gradient Boosting.

Results

Figure 2 reports the performance results by heuristic and algorithmic disambiguation in PENN. The results for simple string matching are shown in circles (String), while those algorithmic disambiguation results are in x-crosses for Logistic Regression (LR), squares for Naïve Bayes (NB), triangles for Random Forest (RF), dashed lines for Support Vector Machine (SVM), and crosses for Gradient Boosting Trees (GB). Full forenames in each dataset are randomly initialized with varying ratios between 0% and 100%. Then, each dataset is randomly divided into equal-sized training and test (validation) sets for machine learning. Clustering outcomes by heuristic and algorithmic disambiguation are evaluated on test sets. The procedure of randomized forename-initialization and data-split is repeated 10 times per full forename ratio, which changes incrementally by 10%. All data points in Figure 2 (a), (b), and (c) represent B-cubed precision, recall, and F1 scores averaged over these 10 iterations per ratio. Standard deviations are not reported as they are negligible (less than 2% of errors).

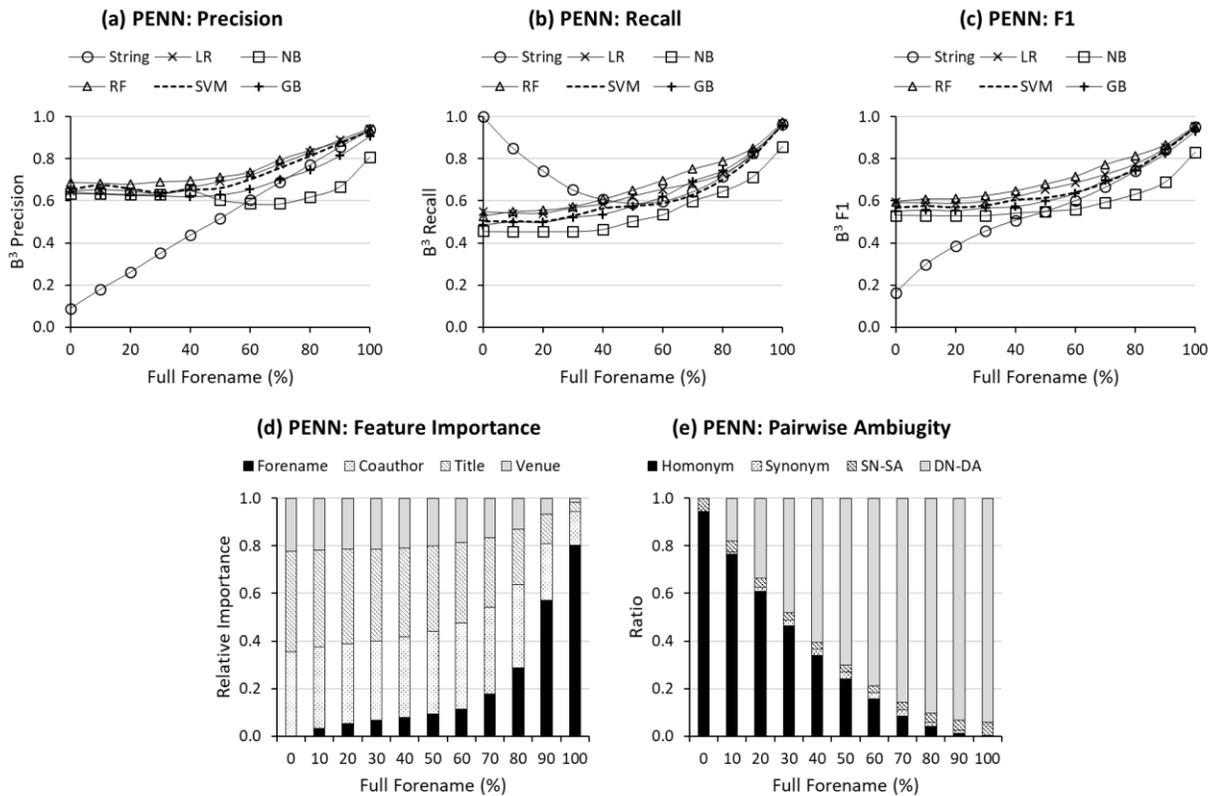

*Figure 2: (a) (b) (c) Performance of Heuristic and Algorithmic Disambiguation, (d) Feature Importance, and (e) Pairwise Ambiguity Types per Full Forename Ratio for PENN*

Heuristic Disambiguation

The circles in Figure 2 represent the performances of heuristic disambiguation in PENN which are evaluated by precision (Figure 2-(a)), recall (Figure 2-(b)), and F1 (Figure 2-(c)). In Figure 2-(a), precision hits below 0.1 when all forenames are initialized ($x = 0\%$), meaning that the heuristic method decides incorrectly name instances to refer to the same author in most cases. In contrast, such an initialization of all forenames leads to an almost perfect recall (0.99 in Figure 2-(b)), meaning that name instances referring to the same authors are almost always found in the same clusters. But this low-precision and high-recall results in a low F1 score (harmonic mean of precision and call) around 0.2 in Figure 2-(c).

As full forename ratios increase, however, mean precision scores by heuristic disambiguation increase almost linearly up to 0.94 in Figure 2-(a). This means that PENN name instances sharing full forenames tend to represent the same author (≈ true-positives) and thus, as more name instances have full forenames, overall precision of heuristic disambiguation also increase. In contrast, recall shows a U-shaped curve in Figure 2-(b). Let's assume that ten exactly same names of an author are being disambiguated. If forenames of five instances are intact and the remaining's forenames are initialized, string matching will divide the truth cluster (10 instances) into two predicted clusters of the same size, comprised of instances that share full forenames and initialized forenames each. If initialized forenames increase from this equilibrium, the size of a cluster where name instances share the same initialized forename will increase, producing higher recall (more true-positives included in the larger cluster) with low precision (more false-positives included, too). This explains the left side of the U-shaped curve. In contrast, if full forenames increase from this equilibrium, the size of a cluster where name instances share the same full forename will increase, producing higher recall but accompanied by high precision (less false-positives). This explains the right side of the U-shaped curve. The U-shaped curve in Figure 2-(b) results from this process happening across blocks in PENN. Importantly, this U-shaped curve in conjunction with the upper-right moving precision plot in Figure 2-(a) indicates that many authors in PENN are distinguishable by their full forename strings: the simple heuristic using full forenames can achieve high mean precision (0.94) and recall (0.95) scores.

Algorithmic Disambiguation

Meanwhile, the performance scores of algorithmic disambiguation in PENN increase as full forenames increase (while other features are the same), across precision, recall, and F1. This is represented by five plots of x-crosses (LR), squares (NB), triangles (RF), dashed lines (SVM), and crosses (GB) moving towards the upper-right corner in Figure 2-(a)/(b)/(c). For example, mean precision scores hover around 0.6~0.7 when all forenames are initialized but rise up to 0.94 (LR, RF) and 0.81 (NB) when all forenames are in a full-string format (Figure 2-(a)). Also, mean recall scores start below 0.6 but increase above 0.95 (LR, RF, SVM, and GB) and 0.86 (NB) as full forename ratios increase (Figure 2-(b)). These observations mean that full forenames are beneficial to algorithmic disambiguation in PENN.

Such results can be better characterized when compared to the performances of heuristic disambiguation. Regarding precision (Figure 2-(a)), especially, machine learning performs better than the heuristic when many forenames are initialized. This indicates that algorithms learn the patterns of true-positive matching effectively using other features such as coauthorship, title, and venue. If most forenames are in a full-string format, however, the heuristic method achieves high precision similar to that achieved by algorithmic disambiguation, even defeating Naïve Bayes after the 60% ratio. This is shown in Figure 2-(a) by the decreasing gaps (*y*-axis) between data points of String (circles) and those of LR (x-crosses), NB (squares), and RF (triangles) over the increasing full forename ratios (*x*-axis). This implies that precision gains by algorithmic disambiguation when compared to the heuristic can be substantial under scarcity of full forenames but become smaller with the increasing full forename ratios in the case of PENN.

Regarding recall, algorithmic disambiguation splits many name instances of an author into different clusters (≈ false-negatives) when initialized forenames are prevalent. As full forenames increase, however, algorithmic disambiguation improves in recall, surpassing the heuristic after the 50% ratio (except Naïve Bayes). This shows that the increasing full forename ratios are also beneficial to algorithms in terms of recall. When precision and recall considered together (Figure 2-(c)), more full forenames are good for both heuristic and algorithmic disambiguation; their performance plots all move towards the upper-right corner. Such a favorable impact becomes more pronounced, especially, for the heuristic method because its performance almost pars the algorithmic methods when more forenames are in a full-string format.

Feature Importance

The aforesaid observations imply that when full forenames are prevalent in PENN, features other than forename may not improve much machine learning performances. To check whether this proposition is true, the importance of each feature is evaluated using Random Forest. Figure 2-(d) reports the evaluation results ($y$-axis) for forename (black), coauthor (dotted), title (diagonal), and venue (grey) across varying full forename ratios ($x$-axis). This evaluation is conducted on the same training datasets for Figure 2, and data points are mean scores.

Figure 2-(d) shows that when all forenames are initialized ($x = 0\%$), the feature importance of forename is zero. This is expected. As matched (positive) and non-matched (negative) name instance pairs all share the same initialized forename in a block, forenames cannot provide any distinctive information. Meanwhile, title (0.42) contributes most to disambiguation performances, followed by coauthor (0.35) and venue (0.22). As the full forename ratios increase, however, forename also increases in importance, surpassing venue, title, and coauthor one by one when the full forename ratios reach 70%, 80%, and 90%, respectively. This confirms the proposition above that the relative contribution of features other than forename decreases when the full forename ratios increase.

Name Ambiguity Type

Figure 2-(e) presents how name ambiguity types in PENN change over the ratios of full forenames. For this, one of four ambiguity types is assigned to a pair of name instances within a block and the ratios of each type over the total of instance pairs are reported after averaged over 10 iterations per full forename ratio. According to the figure, most name instances in PENN are homonymous (black bar) when all forenames are initialized (0%). This is expected because in a block, every instance pairs will have identical initialized forename strings regardless of whether they refer to the same or different authors. As there is no string variation for initialized forenames within a block, synonym and DN-DA types do not exist. As the ratios of full forenames increase, the ratios of SN-SA and DN-DA also increase, while homonymous and synonymous pairs decrease in number. This means that many instance pairs become unambiguous (i.e., if they have identical name strings, they refer to the same author, and if not, they don't) as more forenames are recorded in full format. This also supports the observation above that string-based matching of authors could perform well with increased full forenames.

Cases of KISTI, AMINER, and GESIS

Figure 3 reports disambiguation performance, feature importance, and ambiguity type composition for KISTI. Overall, the results for KISTI are almost the same as those for PENN. Increasing the full forename ratios improves both heuristic and algorithmic disambiguation performances. The feature importance of forenames also increases with larger full forename ratios. The U-shaped recall curve by the heuristic matching shows that full forenames can detect almost all (≈ high recall) and only (≈ high precision) name instances of distinct authors in most cases. Like PENN, as the full forename ratios increase, unambiguous instance pairs (SN-SA and DN-DA) increase, too.

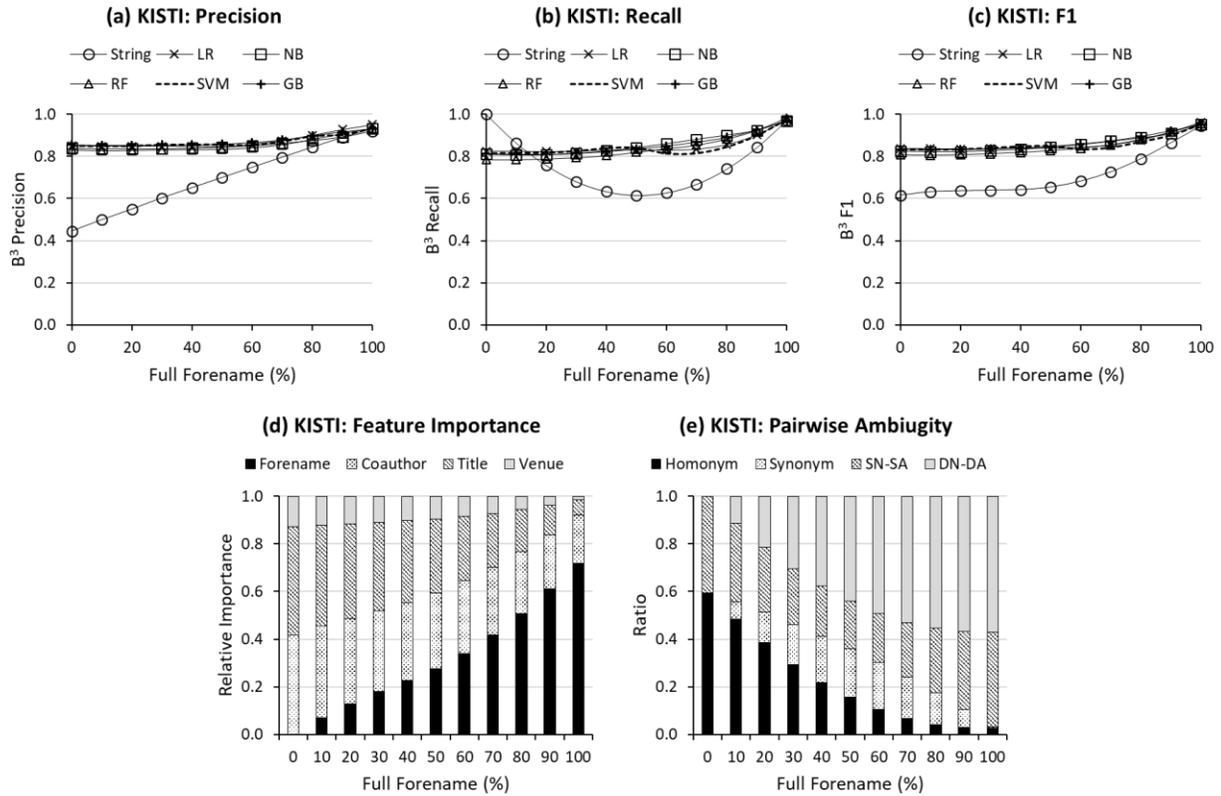

Figure 3: (a) (b) (c) Performance of Heuristic and Algorithmic Disambiguation, (d) Feature Importance, and (e) Pairwise Ambiguity Types per Full Forename Ratio for KISTI

A few differences between KISTI and PENN are worth noting. First, both heuristic and algorithmic disambiguation results show higher mean precision (Figure 3-(a)), recall (Figure 3-(b)), and F1 (Figure 3-(c)) than those for PENN when all forenames are initialized. In addition, the performance gaps between heuristic and algorithm disambiguation are smaller than those for PENN at the 0% of full forename level regarding both precision and recall. These observations indicate that when blocked ($x = 0\%$), name instances in KISTI are less ambiguous than those in PENN. This is confirmed from figure 3-(e) where with all forenames initialized, almost 40% of name instance pairs in a block belong to the ambiguity type of SN-SA, in contrast to the case of PENN where with the 0% full forenames, the SN-SA type makes up less than 5% of all pairs in Figure 2-(e). Regarding feature importance (Figure 3-(d)), forename in KISTI becomes more important than other features at lower full forename ratios than PENN: when compared by bar length, forename surpasses venue in relative importance at 20% (↔ 70% in Figure 2-(d)), title at 60% (↔ 80%), and coauthor at 60% (↔ 90%). This means that full forenames in KISTI are more effective in algorithmic disambiguation than those in PENN.

Unlike PENN and KISTI, however, the results for AMINER in Figure 4-(a) show that increasing full forenames barely improves string matching performances (circles moving flat), meaning that forename string difference does not distinguish most name instances in AMINER. This is expected because AMINER is designed for homonym disambiguation: most name instances in a block share the same name string. So, two name instances may or may not refer to the same author regardless of whether they match on forenames, which makes the string-based matching ineffective in identifying matched instances. This is also supported by the composition of ambiguity types in Figure 4 (e). Even if the ratios of full forename strings increase, the ratios of unambiguous pairs (SN-SA and DN-DA) do not change much, which is contrasted to the dramatic increase of the two types in PENN and KISTI.

Full forename ratios, however, affect the recall of heuristic disambiguation by dividing a truth cluster into two disambiguated clusters (one containing instances with initialized forenames and the other with full forenames), generating the same U-shaped recall curve as one in Figure 2-(b) and 3-(b). Due to this U-shaped recall curve (circles) in Figure 4-(b), the F1 plot (circles) in Figure 4-(c) contains slightly downward curvature although the precision plot (circles) in Figure 4-(a) is almost flat. In relation to the U-shaped curve, Figure 4-(e) shows that as full forenames increase, the ratios of synonyms keeps increase, reaching a peak when the full forename ratio is 50%, and then begin to decrease. This makes sense because at the peak point (50% of full forenames), an author's name instances are likely to be randomized into two halves with full or initialized forenames, creating the largest number of synonymous instance pairs (which leads to the lowest recall in Figure 4-(b)).

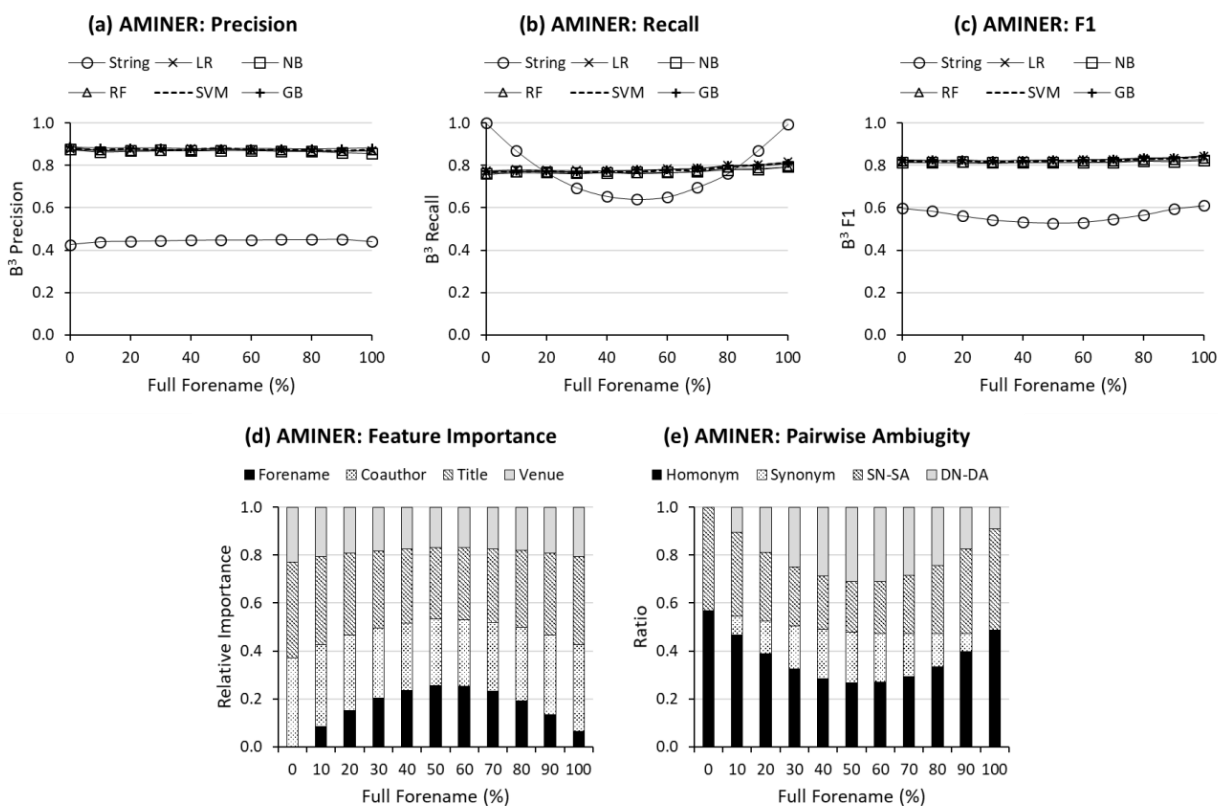

Figure 4: (a) (b) (c) Performance of Heuristic and Algorithmic Disambiguation, (d) Feature Importance, and (e) Pairwise Ambiguity Types per Full Forename Ratio for AMINER

The homonym-centric name composition in AMINER also affects the performances of algorithmic disambiguation: they do not change much over various full forename ratios. This is visualized by plots of LR, NB, RF, SVM, and GB that move horizontally without fluctuation. In addition, the mean precision, recall, and F1 scores of heuristic and algorithmic disambiguation do not converge even when all forenames are in a full-string format, and the plots of algorithmic disambiguation appear mostly above those of the heuristic method. This means that in homonym disambiguation, algorithmic methods can bring substantial contributions to disambiguation results even when full forenames are abundant. A practical implication follows that effectiveness of forename in author name disambiguation should be understood in conjunction with the characteristics of name ambiguity to be solved by a proposed method.

The feature importance evaluation reported in Figure 4-(d) shows that when 50~60% forenames are in a full-string format, forename reaches the highest (0.25) level of importance and decreases regardless of whether the ratios decrease or increase from the 50~60% ratio point. This might be because at these full forename ratios, training instance pairs have the largest variations of similarity over forename, which makes learning by Random Forest more sensitive to them than when the algorithm is trained on pairs with smaller forename similarity variations. Despite such variations, however, forename brings null or small portions of contribution to disambiguation results, which are mostly affected by other features (coauthor, title, and venue) combined.

This observation is also explicable by Figure 4-(e). As full forenames increase, synonymous pairs increase, too. This means that name variants matter for algorithms to learn disambiguation patterns. The combined ratios of ambiguous instance pairs (homonyms + synonyms), however, are almost constant, which seems to set performance bounds of algorithmic disambiguation. This is contrasted to the cases of PENN and KISTI: as the ratios of synonymous pairs change over full forename ratios, the combined ratios of homonyms and synonyms decrease greatly.

In Figure 5, another homonym-centric dataset, GESIS, shows similar patterns as AMINER. Algorithmic disambiguation performances do not change dramatically over full forename ratios, while heuristic disambiguation shows the U-shaped recall performance (Figure 5-(a), (b), and (c)). The feature importance of full forenames is limited, recording below 20% at best (Figure 5-(d)). A major difference between GESIS and AMINER lies in that for GESIS, the ratios of ambiguous name pairs decrease substantially (72% → 45%) with increased full forenames (0% →100%) compared to AMINER in Figure 5-(e). This may explain why both heuristic and algorithmic disambiguation performances improve when full forename strings become more available[13].

---

[13] This study compares instance pairs within a block where they matches on full surnames and initialized first forenames, following the common practice in disambiguation research. Although AMINER and GESIS are created to disambiguate names that match on full names, the blocking also compares name instances with different full forenames if they share the full surname and initialized forenames, generating DN-DA instance pairs in AMINER and GESIS.

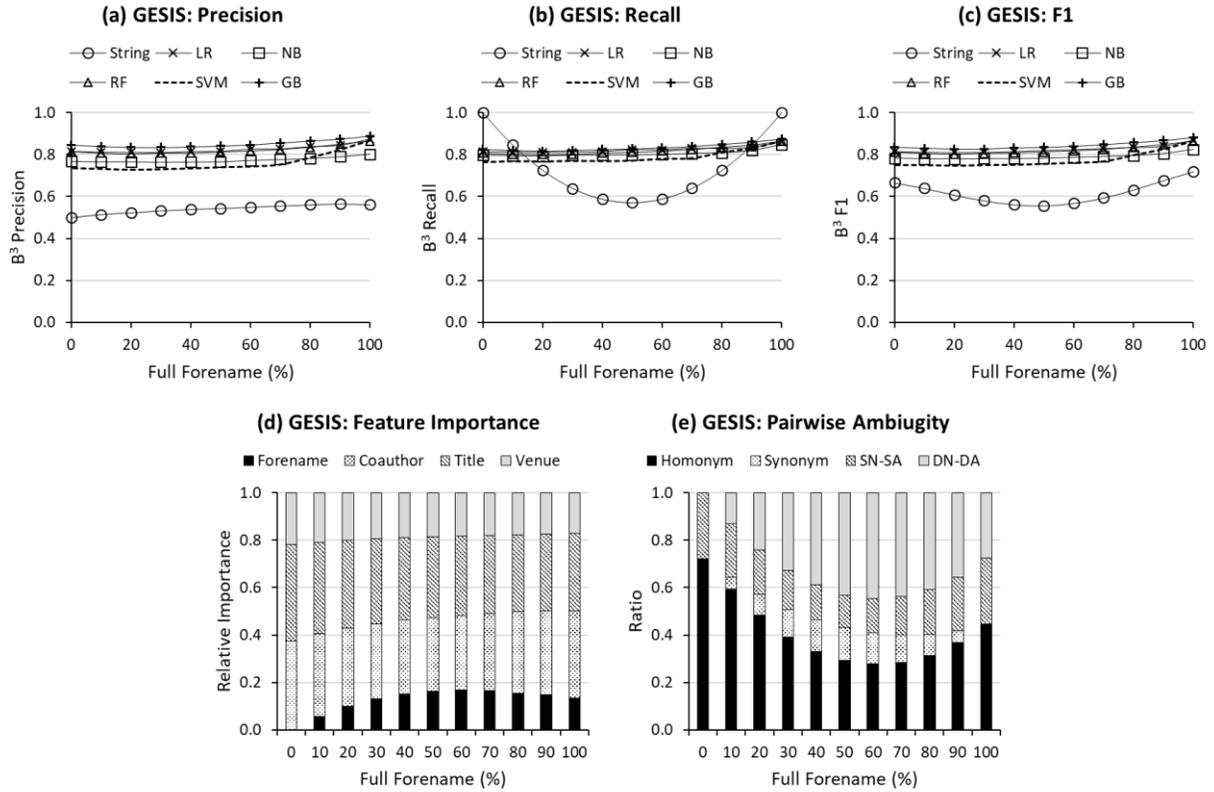

*Figure 5: (a) (b) (c) Performance of Heuristic and Algorithmic Disambiguation, (d) Feature Importance, and (e) Pairwise Ambiguity Types per Full Forename Ratio for GESIS*

Performance Evaluation on Forename *n*-grams

So far, this study shows that full forenames can be effective in disambiguating author names heuristically and algorithmically by varying the full forename ratios in labeled datasets. Another setting of interest is to change the lengths of forename strings to see how incompletely recorded forenames can affect disambiguation performances (e.g., Charles Charlie Brown, Charles Brown, and Charles C. Brown). For this, the length of each forename's characters are counted in each dataset. Non-alphabetical characters (e.g., space, period, apostrophe, and dash) are deleted before counting. Figure 6 reports the cumulative ratios (*y*-axis) of name instances with *n*-gram forename strings (*x*-axis) in each dataset. For example, in AMINER, about 52% (0.5161 on *y*-axis) of name instances have forenames with 5 or less alphabetical characters (max 5-gram on *x*-axis). Half or more name instances have 6 or less characters in PENN, KISTI, and GESIS.

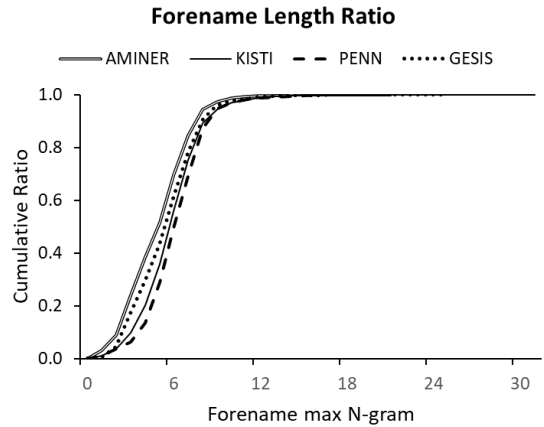

*Figure 6: Cumulative Ratio of Name Instances with Max N-gram Forename*

Next, every forename string in each dataset is truncated into 1 to 10-gram, if available. For example, a name string 'Charles Charlie Brown' is stripped into 'c brown' (1-gram), 'ch brown' (2-gram), 'cha brown' (3-gram) … 'charles cha brown' (10-gram). Then, each dataset is randomly split into training and test sets 10 times per $n$-gram. Figure 6 shows the performance results ($y$-axis) by heuristic (circles) and algorithmic (triangles) disambiguation evaluated on 10 test sets per $n$-gram ($x$-axis) for each dataset. Each row consists of three subfigures reporting mean precision, recall, and F1. For simplicity, only Random Forest results are reported for algorithmic disambiguation. On $x$-axes, 'ALL' means forename strings are used without any truncation, which corresponds to the cases where all available forenames are in full format ($x = 100\%$).

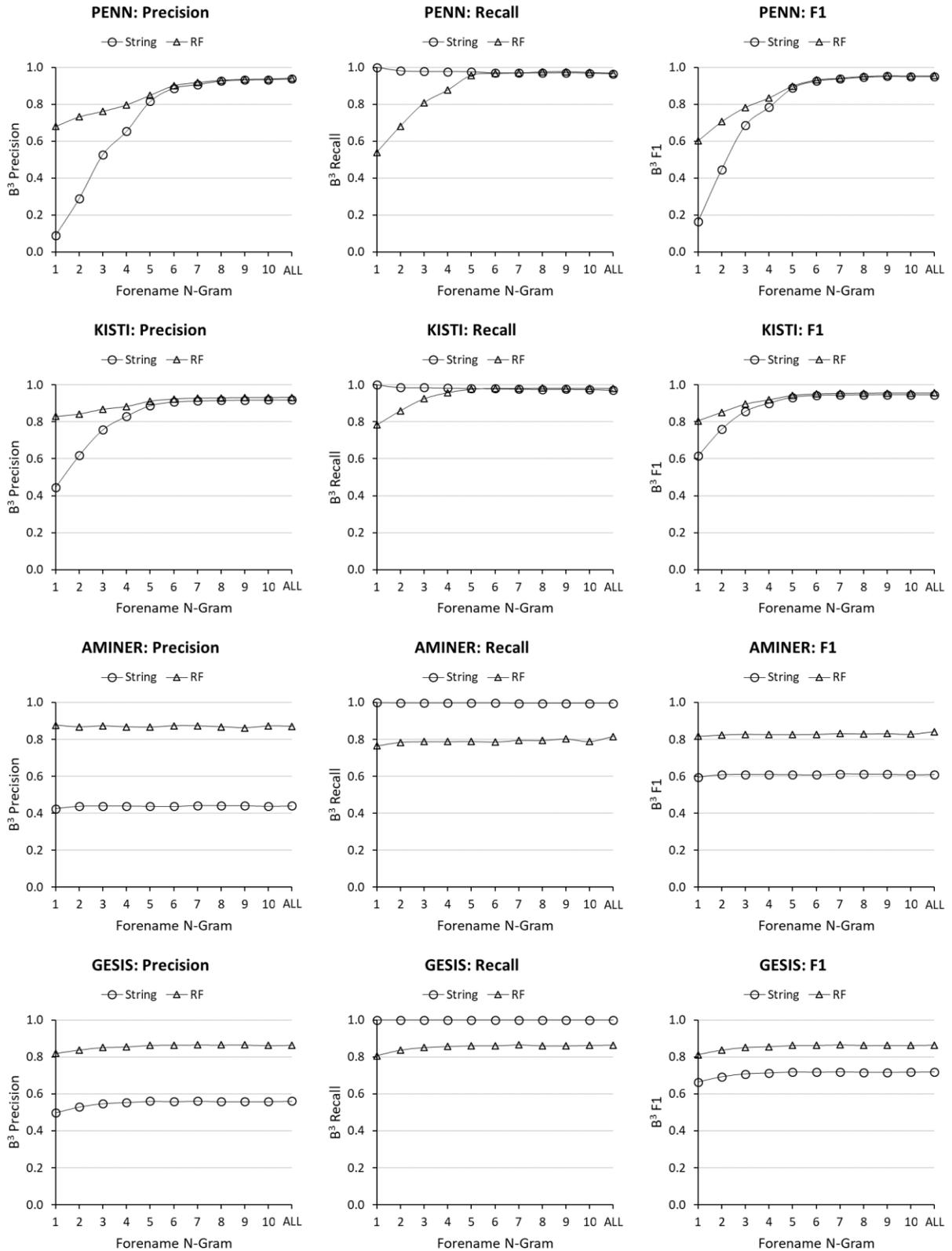

Figure 7: Performance of Heuristic and Algorithmic Disambiguation per Forename N-Gram

According to Figure 7, the performances of both heuristic and algorithmic disambiguation for PENN and KISTI improve as the number of forename characters increases. For example, in subfigures for PENN, mean precision scores by the heuristic method (String) start blow 0.1 with 1-gram, but increase greatly up to 0.93 with 10-gram. Here, the case of all forenames being 1-gram ($x = 1$) corresponds to the case where all forenames are initialized (0%). An exception to this increasing trend by the heuristic disambiguation is its recall results: 1-gram (= all forenames initialized = blocking) produces the highest recall scores, which decrease very slightly as the *n*-gram size increases. This is because small numbers of authors in each dataset are originally recorded in different forename formats; forename matching cannot achieve a perfect recall. Heuristic and algorithmic disambiguation contributes little or slightly to disambiguation results for AMINER and GESIS, regardless of forename lengths. This is in line with the observations in Figure 4 and Figure 5. As most name instances in a block share forenames, truncating their forename strings for an *n*-gram always results in the same situation where most name instances in a block have the same forenames.

The most notable observation in Figure 7 is that the effect of forename strings shows saturation roughly after 5-gram for KISTI and 6-gram for PENN. In other words, the performances of heuristic and algorithmic disambiguation do not improve much with longer forename strings once the maximum lengths of every forename reach 5-gram (KISTI) and 6-gram (PENN). This means that we need part of forename strings to achieve disambiguation performances similar to those obtainable with full-length forenames. Also, note that roughly half of all names instances in each dataset have forenames with 5 or less (KISTI) and 6 or less (PENN) characters (see Figure 6). These two observations imply that if a majority of name instances in KISTI and PENN are recorded correctly within their first 5 or 6 forename characters, the remaining forename characters, however complete they are, do not affect much disambiguation performances. Feature importance evaluation in Figure 8 supports this implication, showing that with the increased lengths, the forename's feature importance increases greatly in PENN and KISTI, but minimally after the saturation point of 5-gram (KISTI) and 6-gram (PENN). In addition, the ratios of ambiguity types in Figure 9 corroborates this finding by showing that after those n-grams, unambiguous name pairs (SN-SA and DN-DA) constitute most of name instance pairs in KISTI and PENN.

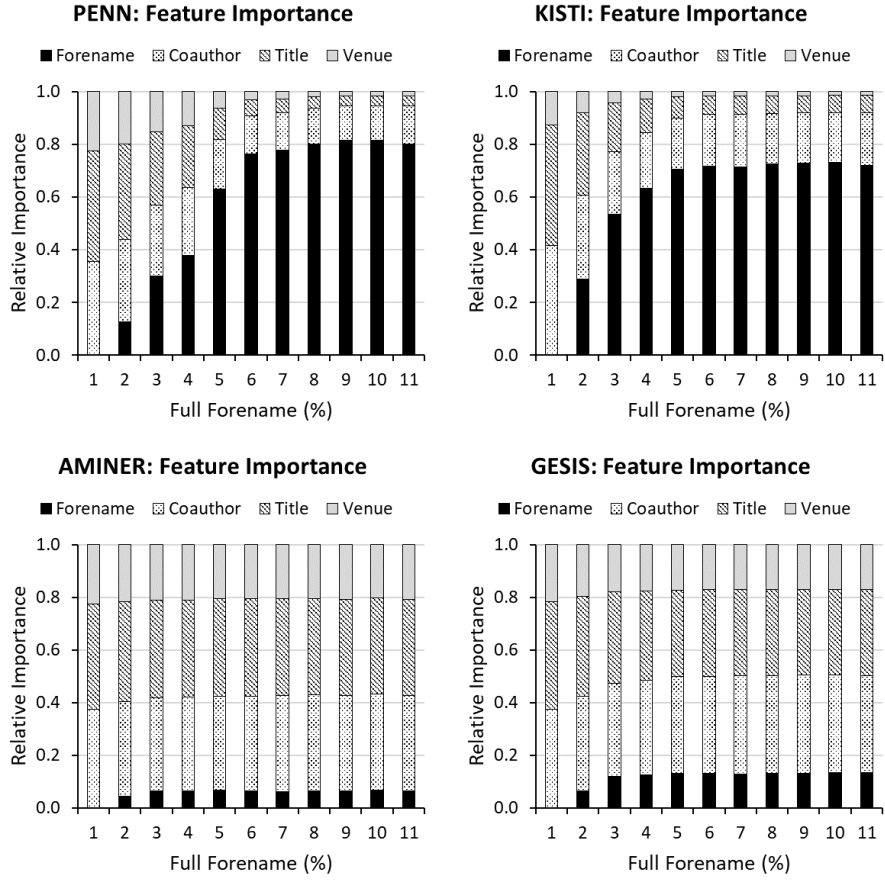

Figure 8: Feature Importance of Algorithmic Disambiguation per Forename N-Gram

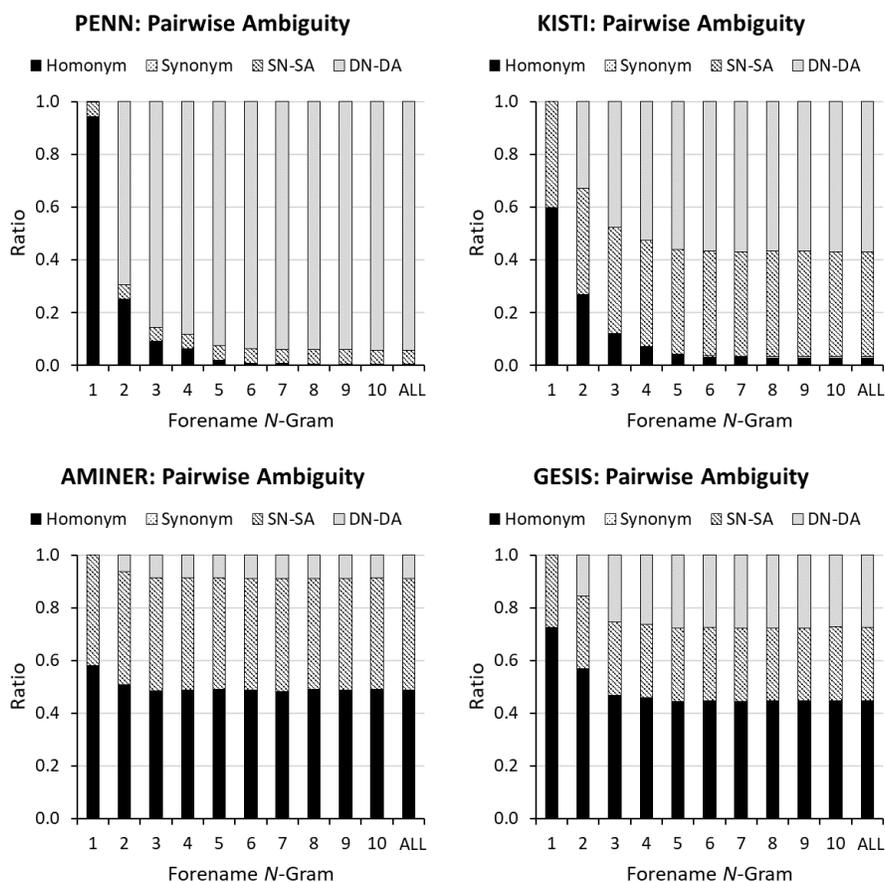

*Figure 9: Ratio of Pairwise Ambiguity Types per Forename N-Gram*

Conclusion and Discussion

This paper evaluated how the ratios of full forenames affect the performances of string-based matching (heuristic) and algorithmic disambiguation methods. Using four labeled datasets (PENN, KISTI, AMINER, and GESIS) in which full forenames were initialized with varying ratios, this paper showed that increasing the full forename ratios improves greatly the performances of both heuristic and algorithmic methods (PENN and KISTI), but hardly does so for the special cases where most name instances in a block have the same forenames (AMINER and GESIS). The performance improvement by full forenames in PENN and KISTI was confirmed by the rising feature importance of forename with increased full forenames. In addition, this study showed that such effectiveness of full forenames is obtainable using 5~6 characters of forename strings.

These findings provide us practical implications about how to improve author name disambiguation. First, as seen in Figure 2~3, name instances can be difficult to disambiguate when their forenames are initialized, but less challenging when full forenames are available for many instances. This implies that disambiguation performance can be enhanced by restoring initialized forenames into a full-string format. For example, MEDLINE name instances in pre-2002 publications in Figure 1 would be disambiguated better if many initialized forenames are replaced by full forenames by, e.g., linking MEDLINE records to external data sources that may contain full forenames. Appendix A demonstrates this potential.

Another implication from Figure 2~3 is that performance gains from algorithmic disambiguation can become less substantial compared to simple string-based matching as full forename ratios increase. This suggests that the impact of forenames on disambiguation results needs to be evaluated before a disambiguation study claims methodological improvements from complicated algorithms or elaborately engineered features. For this, a new disambiguation method may be evaluated under initialized versus full forename settings followed by feature importance assessment or in comparison with results disambiguated by string-based matching. Especially, the latter suggestion supports the idea that string-based matching results need to be baselines in evaluating author name disambiguation (Backes, 2018).

Third, name instances may be blocked based on forename's *n*-gram strings. In Figure 7, using a small portion of forename strings produce decent-to-high precision and recall even by the heuristic method in PENN and KISTI. It implies that the *n*-gram based matching can be used as a blocking method that can achieve good precision and recall, while reducing the size of large blocks which are computationally burdensome (Kim, Sefid, & Giles, 2017). A caveat is, however, that such benefits will be fully realized when many name instances in a target dataset have full forenames. In addition, *n*-gram based blocking will lead to recall loss, asking for an additional high-recall solution.

Most importantly, disambiguation studies need to properly characterize name ambiguity in data which they attempt to disambiguate. For datasets like PENN and KISTI in which authors are distinguishable by full forenames in many cases, sophisticated string-matching techniques may produce decent performances although they cannot defeat feature-based machine learning. In contrast, for others like AMINER and GESIS in which distinct authors tend to have the same full forenames (homonyms), relying heavily on name strings can lead to inaccurate disambiguation, while algorithmic disambiguation can contribute substantially to the task. For this, more efforts are required to study name ambiguity itself, which can be quite challenging as reported in Ackermann and Reitz (2018) to detect homonym cases in a digital library.

A few limitations of this study are worth noting to direct future studies on this topic. First, although this study addressed the synonym case where an author is represented by initialized and full forenames or by varying lengths of forenames, other types of synonyms (e.g., different by edit-distance, flipped ordering of forename tokens, etc.) are not considered. Such synonym cases are reported to be infrequent (Torvik & Smalheiser, 2009), but their impact on name disambiguation may not be negligible for certain groups of authors whose names are susceptible to such variations (e.g., German, Hispanic, and Slavic names). Second, this study implemented the commonly used hierarchical agglomerative clustering for which the similarity distance among instances is predicted by popular classifiers. As a heuristic disambiguation method, name instances that match on all available forename strings were assumed to refer to the same author. However, there are a variety of string-based matching methods. The evaluation results reported in this study may be different if our datasets are evaluated by different algorithmic and heuristic methods. To assist validation and comparative studies, the datasets that are split into training and test sets with varying forename ratios and lengths are available upon request.

Despite the limitations, this study provides empirical findings suggesting that author name disambiguation studies need to consider the effect of forenames on disambiguation performances. Depending on the magnitude of full forename's impact, major methodological changes may follow, as illustrated above. These changes are expected to improve practically disambiguation performances in evolving digital libraries in which author names with full forenames keep increasing in number and are posing new challenges to efforts trying to disambiguate author names at scale.

Acknowledgement

This work was supported by grants from the National Science Foundation (#1561687 and #1535370), the Alfred P. Sloan Foundation and the Ewing Marion Kauffman Foundation.

Appendix A

This section demonstrates how initialized forenames can be restored into full ones via record linkage and how such restoration can affect performances for both algorithmic and heuristic disambiguation. First, 543,205 paper records in MEDLINE for the 1992~2001 (10 years) period were selected and matched through title words (lower-cased and ASCII-converted) to publication records in Microsoft Academic Graph (MAG). After excluding matching results for short titles (less than 5 words) and duplicates, a total of 395,466 (72.80%) MEDLINE records were linked to MAG. Among 1,584,023 name instances (95.75% have initialized forenames) in the MEDLINE-MAG linked data, 85.52% of them were restored from initialized forenames into full ones.

Second, the 1.58M instances were matched to author profiles in ORCID data[14], resulting in 100,383 ORCID-linked instances. Among them, 3,188 ambiguous name instances of 236 blocks (associated with 5 or more unique ORCIDs in the original MEDLINE records) were selected as labeled data. The data were randomly split 10 times into training and test datasets for before- and after-restoration situations, respectively. Then, the same heuristic and algorithmic disambiguation procedures described in Methods were applied. Figure 10 shows disambiguation performances of the string-based matching heuristic (HEUR) and the Random Forest (ALGO) tested on datasets with initialized forenames (INI) and restored ones (FULL).

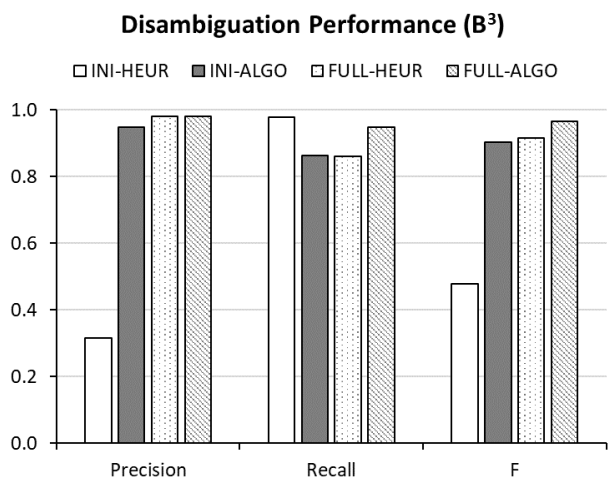

*Figure 10: Disambiguation Performance of Heuristic and Algorithmic Disambiguation on MEDLINE-ORCID Linked Labeled Data*

According to the figure, adding full forenames improves substantially the performances of both heuristic and algorithmic disambiguation. Specifically, the heuristic matching produced mean $B^3$-F score increased from 0.48 (INI-HEUR) to 0.92 (FULL-HEUR) with added full forenames. The algorithmic disambiguation produced mean $B^3$-F score improved from 0.90 (INI-ALGO) to 0.97 (FULL-ALGO).

---

[14] https://figshare.com/articles/ORCID_Public_Data_File_2018/7234028

Appendix B

This section explores how feature similarity measures can affect the performances of an algorithmic disambiguation. The cosine similarity of n-gram's Term-Frequency was used over all features in this paper (All N-Gram). For comparison, two combinations of commonly used similarity measures were tested. Distinct1 consists of n-gram's TF cosine (author name), Jaro-Winkler (coauthorship), and token-based cosine (title and venue). Distinct2 consists of Jaro-Winkler (author name and coauthorship) and token-based Jaccard (title and venue). Figure 11 and Figure 12 report the disambiguation performance of Random Forest and feature importance assessed for the three similarity calculation methods (All N-Gram, Distinct1, and Distinct2) applied to KISTI and AMINER.

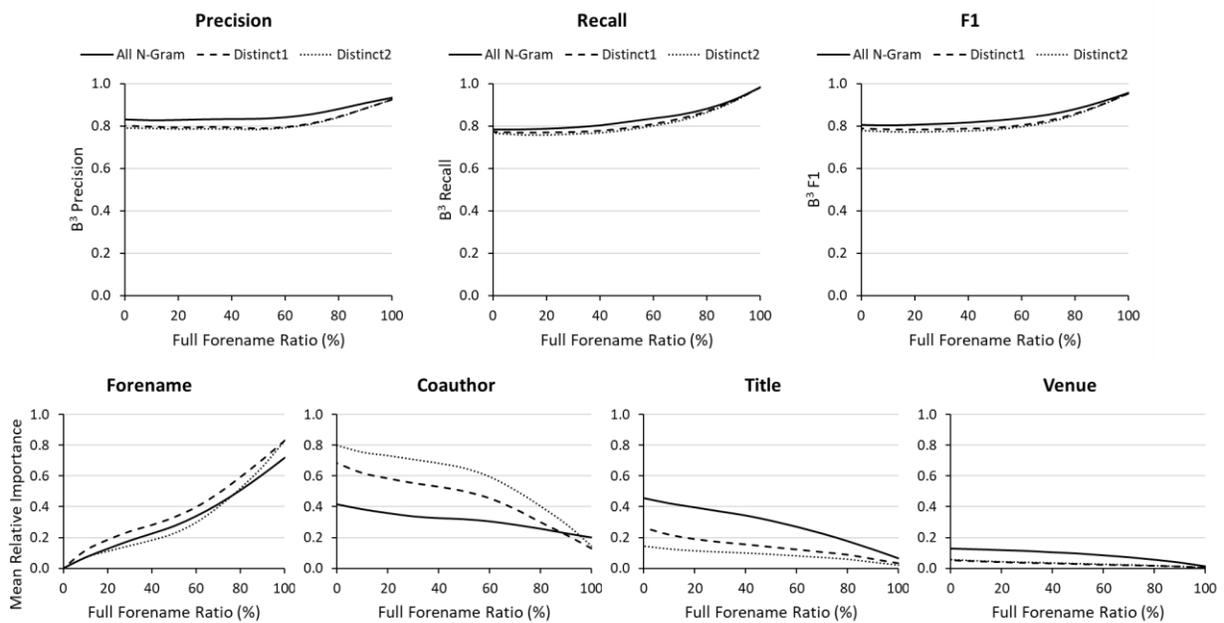

*Figure 11: Disambiguation Performance and Feature Importance of Three Similarity Calculation Methods for KISTI*

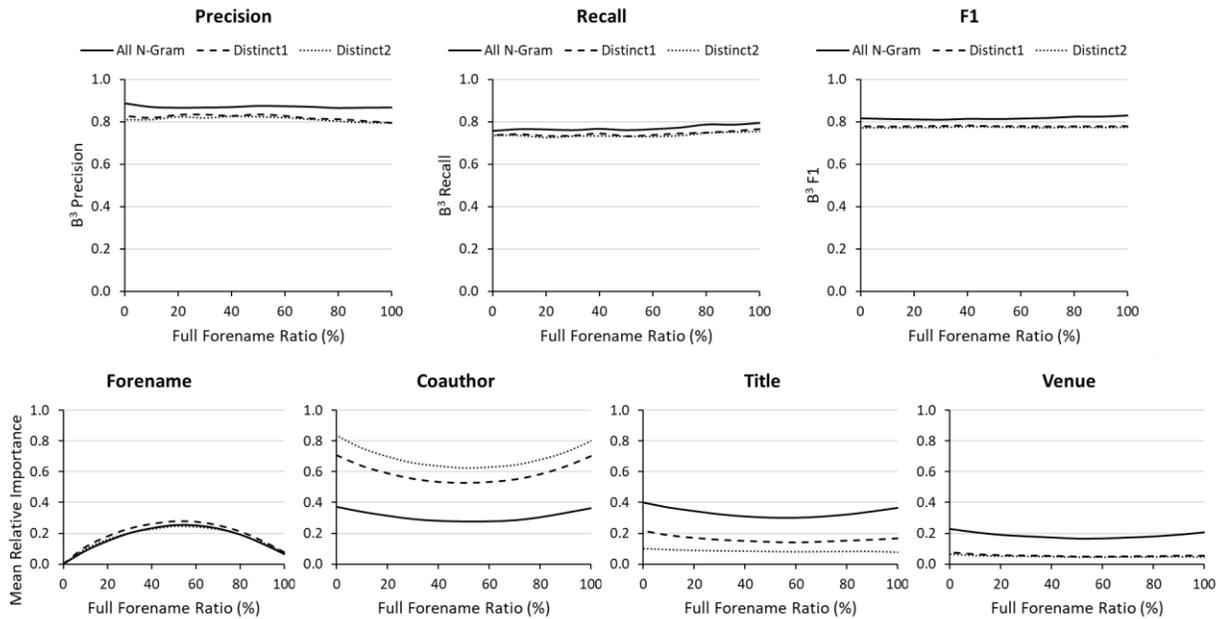

*Figure 12: Disambiguation Performance and Feature Importance of Three Similarity Calculation Methods for AMINER*

According to the figures, the feature similarity calculation approach of this paper (solid line) performs slightly better than other two alternatives (dashed and dotted lines). While the contributions of forenames do not change much per method, those of coauthorship increase a lot when Distinct1 and Distinct2 are applied. But such an increase comes at the decrease of contributions by title and venue.